\pdfoutput=1
\documentclass[11pt,a4paper]{article}
\usepackage{amsmath,amssymb}
\usepackage{braket}
\usepackage{url}
\usepackage{mathtools}
\usepackage{amsmath}              
\usepackage{graphics,graphicx,epsfig,ulem} 
\usepackage{subfigure,subfig}
\usepackage{feynmp}
\usepackage{slashed}

\DeclareMathAlphabet{\mathbold}{U}{zeur}{b}{n}

\renewcommand\[{\left[}
\renewcommand\]{\right]}

\def\beq{\begin{equation}}
\def\eeq{\end{equation}}
\def\[{\begin{equation}}
\def\]{\end{equation}}

\begin{document}
\numberwithin{equation}{section}

\title{
{\normalsize  \mbox{}\hfill IPPP/15/18, DCPT/15/36}\\
\vspace{2.5cm}
\Large{\textbf{Diagrammatic computation of multi-Higgs processes at\\ very high energies: scaling $F_{\bf holy\,grail}$ with MadGraph}}}

\author{Valentin V. Khoze\\[4ex]
  \small{\it Institute for Particle Physics Phenomenology, Department of Physics} \\
  \small{\it Durham University, Durham DH1 3LE, United Kingdom}\\[0.2ex]
  \small{\tt valya.khoze@durham.ac.uk}\\[0.8ex]
}

\date{}
\maketitle

\begin{abstract}
 \noindent At very high energies scattering amplitudes in a spontaneously broken gauge theory into multi-particle final states 
are known to grow factorially with the number of particles produced. Using simple scalar field theory models with and without the VEV, 
we compute total cross-sections with up to 7 particles in the final state at the leading order in perturbation theory 
 with MadGraph. By exploring the known scaling properties of the multi-particle rates with the number of particles,
we determine from these the general $n$-point cross-sections in the large-$n$ limit. In the high-multiplicity regime we are considering, 
$n\gg1$ and $\lambda n$=fixed, the perturbation theory becomes strongly coupled with the higher-order loop effects contributing 
increasing powers of $\lambda n$. In the approximation where only the leading loop effects are included, 
we show that the corresponding perturbative cross-sections grow exponentially and ultimately violate perturbative unitarity.
This occurs at surprisingly low energy scales $\sim 40-50$ TeV with multiplicities above $\sim 150$. It is expected that a
repair mechanism or an extension of the theory has to set-in before these scales are reached, possibly involving
a novel non-perturbative dynamics in the a priori weakly coupled theory.
 \end{abstract}

\bigskip
\thispagestyle{empty}
\setcounter{page}{0}

\newpage


\section{Introduction}\label{sec:intro}

We are interested in scattering processes at very high energies into $n$-particle final states in the limit $n\gg 1$.
In this case the well-known problem of divergences of  {\it large} orders of perturbation theory
 \cite{Dyson,Lipatov:1976ny,Brezin:1976vw,'tHooft:1977am}, is realised instead at the {\it leading} order.
 This is because even the leading-order Born diagrams for the $n$-point scattering amplitudes are expressed in terms  of
Feynman diagrams with large numbers of vertices, and with the numbers of diagrams growing factorially with $n$.
At sufficiently high energies the production of high multiplicity final states, with $n$ greater than the inverse coupling constant,
is kinematically allowed 
and the $n$-point scattering amplitudes near the multi-particle mass thresholds grow as $n!$ at leading order -- i.e. tree level in a weakly coupled theory.

In the simplest scenarios, production rates for such final states can be considered in a quantum field theory with a single scalar 
field of mass $M$ and the coupling $\lambda$. The model with a non-vanishing VEV $\langle h \rangle = v$,
\[
{\cal L}^{\rm SSB} \,= \, \frac{1}{2}\, \partial^\mu h \, \partial_\mu h\, -\,  \frac{\lambda}{4} \left( h^2 - v^2\right)^2
\,,
\label{eq:LSSB}
\]
is a simplified version of the Higgs sector of the SM in the unitary gauge, describing neutral Higgs bosons of mass 
$M = \sqrt{2\lambda}\,v$.   We will refer to this model as the 
theory with spontaneous symmetry breaking (SSB) and it will be our principal case of interest for this paper.
In addition to \eqref{eq:LSSB} will also consider multi-particle amplitudes in an even simpler
$\phi^4$ theory, with no spontaneous symmetry breaking,
\[
{\cal L}^{\rm no\, SSB} \,= \, \frac{1}{2} \left(\partial \phi\right)^2 - \frac{1}{2} M^2 \phi^2 - \frac{1}{4} \lambda \phi^4
 \,.
\label{eq:LnoSSB}
\]
The model \eqref{eq:LnoSSB} with an unbroken ${\cal Z}_2$  symmetry has been widely used in computations of multi-particle rates in the 1990's as
reviewed in Ref.~\cite{LRT} and other papers referred therein and below.

\bigskip

 Our goal here is to compute the multi-particle rates directly in perturbation theory using
 one of the current state of the art publicly available numerical techniques, in the current case -- MadGraph 5 \cite{Alwall:2011uj}.
 The continuation procedure from moderate values of $n=7$ particles in the final state, where our calculations are performed,
 to the regime with $n \sim 10^2 - 10^3$ will be set-up and carried out in Section {\bf \ref{sec:3}} based on the known scaling properties of the multi-particle cross-sections with $n$, as will be outlined next in Section {\bf \ref{sec:2}}.

\medskip
\section{Multi-particle production rates and the holy grail function}
\label{sec:2}
\medskip

Let us consider the multi-particle limit $n\gg 1$ for the $n$-particle final states, and scale the energy $\sqrt{s} = E$ linearly with $n$,  $E \propto n$, 
keeping the coupling constant small at the same time, $\lambda \propto 1/n.$ 
It was first argued in \cite{LRST} (for a review of subsequent developments see \cite{LRT}) that in this double-scaling limit the production cross-sections $\sigma_n$ have a characteristic exponential form,
\[
\sigma_n \,\sim\, e^{ \,n F(\lambda n, \,\varepsilon)} \,,
\quad {\rm for}\,\, n \to \infty\,,\,\, \lambda n ={\rm fixed}\,,\,\, \varepsilon ={\rm fixed}\,,
\label{eq:hg}
\]
where $\varepsilon$ is the average kinetic energy per particle per mass in the final state,
\[
\varepsilon \,=\, (E-nM)/(nM)\,,
\label{eq:eps}
\]
and $F(\lambda n, \varepsilon)$ is a certain a priori unknown function of two arguments.  $F(\lambda n, \varepsilon)$  is often referred to 
as the `holy grail' function for the multi-particle production\footnote{Equation~\eqref{eq:hg} can be equivalently written in the form \cite{LRST}
$\sigma_n \,\sim\, \exp \left[ {\lambda}^{-1}\, {\cal F}(\lambda n, \epsilon) \right]$, using the rescaling  
$ {\cal F}(\lambda n, \epsilon) = \lambda n F(\lambda n, \varepsilon)$, which points towards a semi-classical interpretation 
of the rate in the $\lambda \to 0$ limit.} in the perturbative sector.

\medskip

At small values of $\epsilon$ and $\lambda n$, the large-$n$ behaviour in \eqref{eq:hg} has been verified explicitly and the function 
$ F(\lambda n, \epsilon)$ was computed in \cite{LRST} for the VEV-less scalar theory \eqref{eq:LnoSSB}, and later
in \cite{VVK2} for the theory with the VEV \eqref{eq:LnoSSB} and more generally in a Gauge-Higgs theory. These computations were carried out in perturbation theory at tree-level combined with the simplifications arising in the non-relativistic limit
$\varepsilon \ll 1$ for the final state particles. This approach has allowed for the analytic  derivation the corresponding tree-level amplitudes and their phase-space integration for all values of $n\gg 1$. It was found that the dependence of the holy grail function on
its two arguments, $\lambda n$ and $\varepsilon$, factorises into individual functions of each argument
\[
\log \sigma_n^{\rm tree}|_{n\to \infty}\to\, n F^{\rm tree}(\lambda n, \,\varepsilon) \,=\, n\left(f_0(\lambda n)\,+\, f(\varepsilon)\right)\,,
\label{eq:nFtree}
\]
and the two independent functions are given by the following expressions in the model \eqref{eq:LnoSSB} Ref.~\cite{LRST}:
\begin{eqnarray}
\label{f0noSSB}
f_0(\lambda n)^{\rm no\, SSB}&=& \log\left(\frac{\lambda n}{16}\right) -1\,, \quad n={\rm odd}\,,
\\
\label{fenoSSB}
f(\varepsilon)^{\rm no\, SSB}|_{\varepsilon\to 0}&\to& f(\varepsilon)_{\rm asympt}\,=\, 
\frac{3}{2}\left(\log\left(\frac{\varepsilon}{3\pi}\right) +1\right) -\frac{17}{12}\,\varepsilon\,,
\end{eqnarray}
and in the Higgs model \eqref{eq:LSSB} Ref.~\cite{VVK2} respectively:
\begin{eqnarray}
\label{f0SSB}
f_0(\lambda n)^{\rm SSB}&=&  \log\left(\frac{\lambda n}{4}\right) -1\,, 
\\
\label{feSSB}
f(\varepsilon)^{\rm SSB}|_{\varepsilon\to 0}&\to& f(\varepsilon)_{\rm asympt}\,=\, 
\frac{3}{2}\left(\log\left(\frac{\varepsilon}{3\pi}\right) +1\right) -\frac{25}{12}\,\varepsilon\,.
\end{eqnarray}
These results arise from integrating the known expressions \cite{LRST,VVK2}
for the tree-level amplitudes near the multi-particle thresholds,
\begin{eqnarray}
\label{AnoSSB}
{\cal A}_{1^*\to n}^{\rm no\,SSB} &=&n!\,  \left(\frac{\lambda}{8M^2}\right)^{\frac{n-1}{2}}\exp\left[-\frac{5}{6}\,n\, \varepsilon \right]\,,
\\
\label{ASSB}
{\cal A}_{1^*\to n}^{\rm SSB} &=&n!\,  \left(\frac{\lambda}{2M^2}\right)^{\frac{n-1}{2}}\exp\left[-\frac{7}{6}\,n\, \varepsilon \right]\,,
\end{eqnarray}
over the Lorentz-invariant phase-space, 
$\sigma_n =\frac{1}{n!} \int \Phi_n  \, \left|{\cal A}_{n}\right|^2$, in the large-$n$ non-relativistic approximation.
In particular, the ubiquitous factorial growth of the large-$n$ amplitudes in \eqref{AnoSSB}-\eqref{ASSB} translates into the 
 $\frac{1}{n!} |{\cal A}_{n}|^2 \sim n! \lambda^n \sim e^{n\log(\lambda n)}$ factor in the cross-section, which determines the 
 function $f_0(\lambda n)$ in \eqref{f0noSSB} and \eqref{f0SSB}. The energy-dependence of the cross-section is dictated by
 $f(\varepsilon)$ in Eq.~\eqref{eq:hg}, and this function arises from integrating the $\varepsilon$-dependent factors in \eqref{AnoSSB}-\eqref{ASSB}
 over the phase-space, giving rise to the small-$\varepsilon$ asymptotics in \eqref{fenoSSB},\eqref{feSSB}.
 
 \medskip
 
 An important for our forthcoming analysis point to make, is that the separability or factorisation of the $\lambda n$-
 from the $\varepsilon$-dependence on the right hand side of Eq.~\eqref{eq:hg}, is the general consequence of the
 tree-level approach, i.e. it does not require taking the non-relativistic limit $\varepsilon \ll 1$ .
 This is because the entire $\lambda$-dependence of the full tree-level result $\sigma_n \propto \lambda^n$ is contained in the
 $f_0$ function. Hence, given that the dependence on $n$ could enter $f(\varepsilon)$ only in the combination  $\lambda n$ and that the dependence 
 on $\lambda$ is already fully accounted for,\footnote{We recall that this argument applies at tree level.}
  the function $f(\varepsilon)$ does not depend on $n$. We thus should be able to 
 determine $f(\varepsilon)$ from the fixed-$n$ direct calculations of the cross-sections, and using the large-$n$ scaling 
 arguments suggested by \eqref{eq:hg}. In section {\bf \ref{sec:3}} we will proceed to construct
 $f(\varepsilon)$ from the cross-sections data at $n=7$ which will be computed numerically using MadGraph \cite{Alwall:2011uj}.
 
 It should also be kept in mind that the holy grail function in the cross-section formula Eq.~\eqref{eq:hg} contains not only
 the tree-level contributions but also the loop contributions with an arbitrary number of loops, which to a large extend give the dominant contributions for $\lambda n \gtrsim 1$, as will be explored in more detail below.

\medskip
\section{Results}
\label{sec:3}
\medskip

 \begin{figure}[t]
\begin{center}
\begin{tabular}{cc}
\hspace{-0.5cm}
\includegraphics[width=0.4\textwidth]{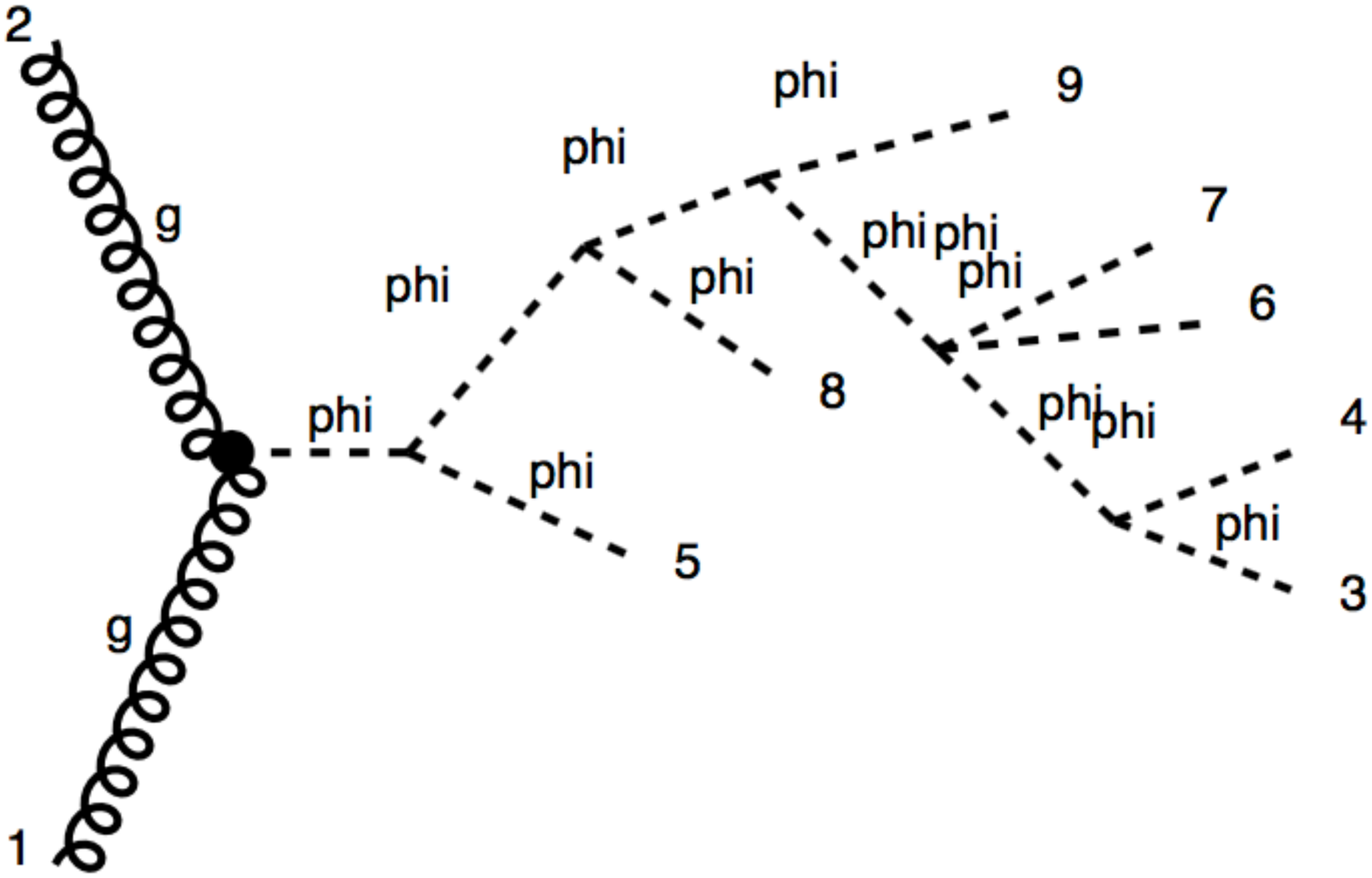}
\hspace{.1cm}
&
\hspace{.1cm}
\includegraphics[width=0.4\textwidth]{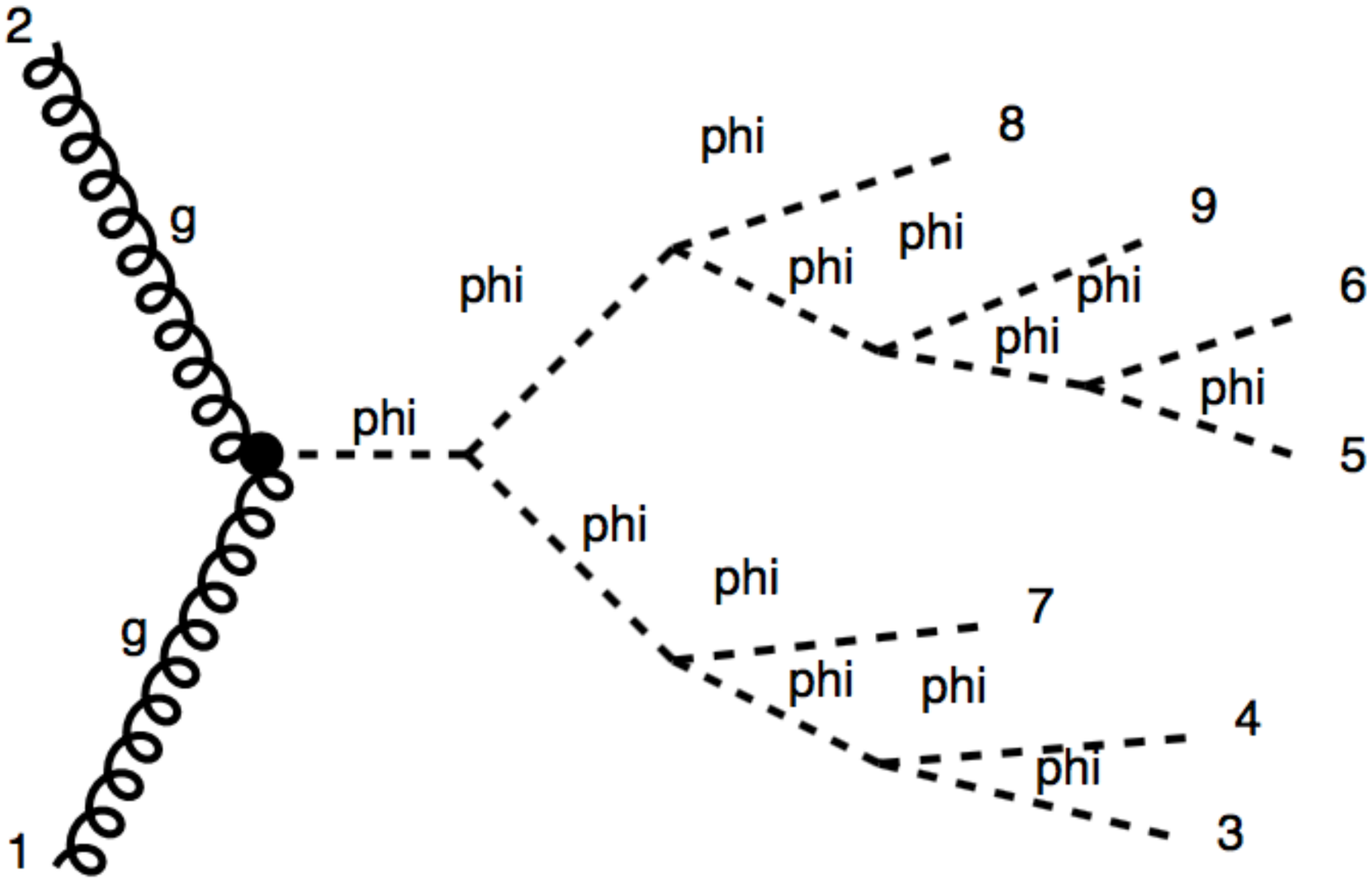}
\\
\end{tabular}
\end{center}
\vskip-1cm
\caption{
Representative Feynman diagrams for the $g\,g\,\to\,   7\, h$ process in the SSB theory \eqref{eq:LSSB}.
}
\label{fig:f1}
\end{figure}

With MadGraph 5 \cite{Alwall:2011uj} we can compute quite efficiently total cross-sections for scattering processes with 
$2\to 7$ particles. As we are mainly interested in producing the multi-particle final state we can make certain simplifications 
with respect to the characterisation of the initial 2-particle state. First, we take that the scattering process proceeds originates from the
gluon fusion in the initial state, producing the highly-virtual single higgs boson $h^*$ via the effective  $ggh$ vertex 
$\frac{\alpha_s}{12\pi v} \, h \,{\rm tr} G^{\mu\nu}G_{\mu\nu}$, which is followed by the $1^* \to n$ process
computed in the scalar theory \eqref{eq:LSSB} or \eqref{eq:LnoSSB},
\[ 
g\,g\,\to\,  h^* \, \to\, n\times h\,.
\label{eq:singleh}
\]
In principle, as is well-known, the use of the point-like effective $ggh$ coupling approximation is not justified at high energies for producing 
realistic cross-sections. Finite top mass effects in loop will result in the form-factor in front of the exponential factor in the cross-section.
However, in our case the single effective vertex is only a gimmick -- in practice we will be computing {\it ratios} of the cross-sections at the same values of
 energy for different $n$. These ratios are insensitive to the bad high-energy behaviour of the effective vertex in the initial state. 
In our final results plotted in Fig.~\ref{fig:f6} we will include the effect of the Higgs form-factor, as shown in Eq.~\eqref{eq:sigmaff}.

 \begin{figure}[t]
\begin{center}
\includegraphics[width=0.7\textwidth]{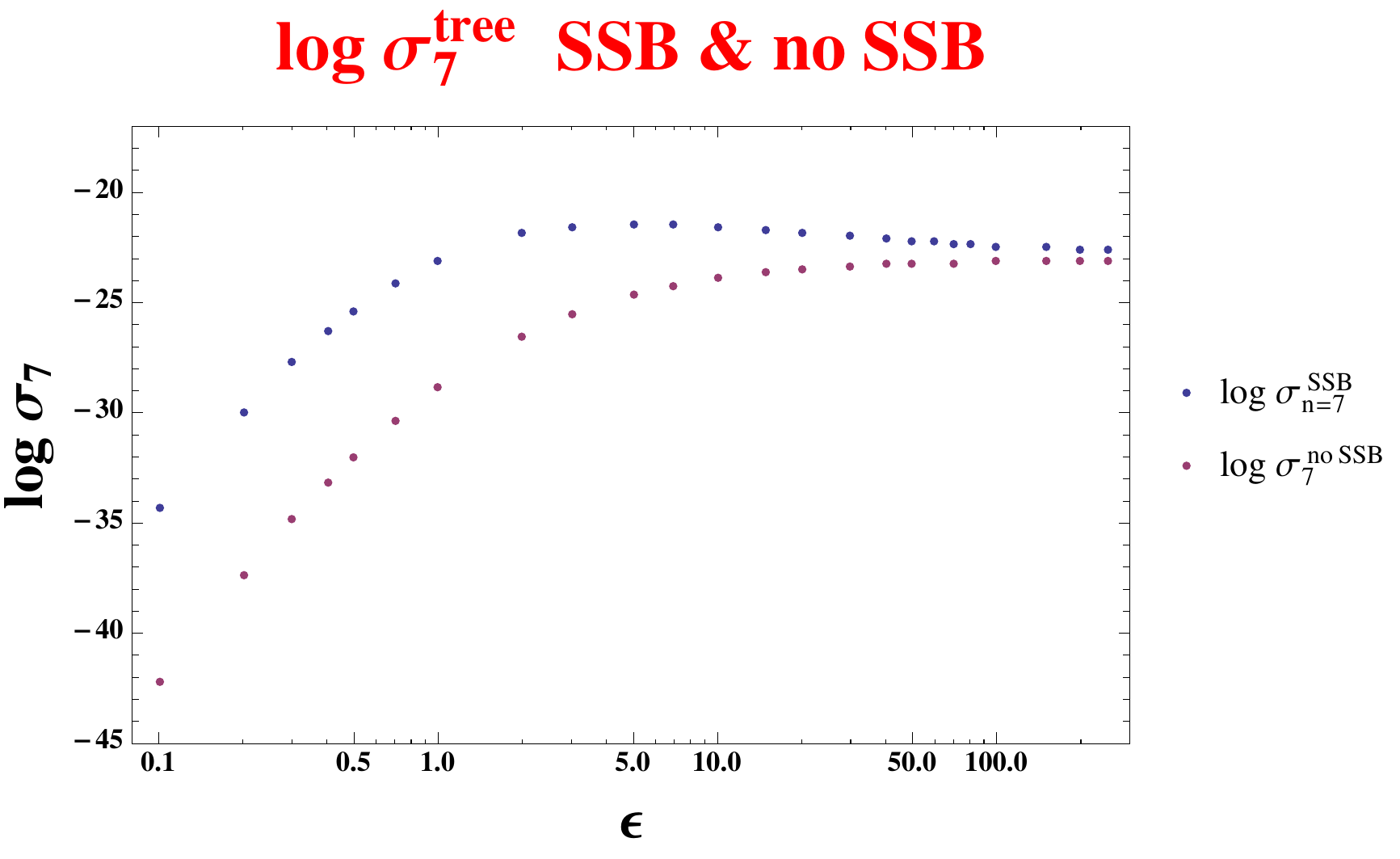}
\end{center}
\vskip-0.5cm
\caption{
Selection of data points for  $\log \sigma_7$ as the function of $\varepsilon$ computed with MadGraph
in the models with and without spontaneous symmetry breaking (the upper and the lower contours).
}
\label{fig:f2}
\end{figure}

At $n=7$ in the Higgs theory with SSB \eqref{eq:LSSB} MadGraph computes 34,300 diagrams 
(two of which are shown in Fig.~\ref{fig:f1}) contributing to the tree-level 
scattering amplitude. The cross-section values $\sigma_7^{\rm SSB}$ were then computed for different energies
on a grid of $\sim 30$ points with values of $\varepsilon =E/(7M)-1$ ranging from 0.001 (nearly at the mult-iparticle threshold)
to 250 (ultra-relativistic final state). We have chosen $M=125$ GeV and set $\lambda=1$\footnote{This can -- and for the applications in plots in Figs.~\ref{fig:f5},
\ref{fig:f6} will be -- rescaled to
the physical value $\lambda \simeq 1/8$ using the fact that for each $n$,  $\sigma_n^{\rm tree} \propto \lambda^{n-1}$.}. 
To give an example, at $\varepsilon=1$ which corresponds to 1750 GeV, the rate is 
$\sigma_7^{\rm SSB}=\, 8.913 \times 10^{-11} \, \pm 2.74\times 10^{-13}  \, {\rm pb}$, and at
$\varepsilon=30$ corresponding to 27125 GeV, the rate is 
$\sigma_7^{\rm SSB}=\, 2.818 \times 10^{-10}  \, \pm 9.02\times 10^{-13}  \, {\rm pb}$. 
More data points for $\sigma_7^{\rm SSB}$ for $0.1 \le \varepsilon\le 250$ are shown in Fig.~\ref{fig:f2} as the upper contour (in blue).

\subsection{Extracting $f(\varepsilon)$ from computed cross-sections}

The expression in \eqref{eq:nFtree} contains only the contributions growing with $n$; it does not include subleading corrections, and
 is valid, as it stands, only in the $n \to \infty$ limit. To be able to work at moderately large values of $n$, such as 
$n=6,7$, we now generalise this by including the sub-leading corrections in $n$. In general, they are of the form ${\cal O}(\log n)$  and ${\cal O}(n^0)$,
so that in total we have,
\[
\log \sigma_n^{\rm tree} \,=\, n\left(f_0(\lambda n)\,+\, f(\varepsilon)\right) \,+\, c_0\log n  \,+\, c_1 f_1(\varepsilon) \,+\, c_2 \,+\, {\cal O}(1/n)\,.
\label{eq:nFtree_imp}
\]
Here $c_0$, $c_1$ and $c_2$ are some unknown constants, and $f_1(\varepsilon)$ is a new function of the kinetic energy.
For example, by carrying out the phase-space integration of the non-relativistic amplitudes in the small-$\varepsilon$ limit
beyond the leading order in $n$ one finds,
\[
\log \sigma_n^{\rm tree} \,\to \, n \left( f_0(\lambda n)\,+\, (3/2) \log \varepsilon\right)\,+\, c_0\log n  \,-\, (5/2) \log \varepsilon \,+\, c_2 \,+\, {\cal O}(1/n)\,.
\nonumber
\]
Returning to Eq.~\eqref{eq:nFtree_imp} we now consider the difference between the rates at $n$ and $n-1$. This
allows us to extract $f(\varepsilon)$ directly from the $\log \sigma_n \,-\, \log \sigma_{n-1} $ data as follows:
\[
f(\varepsilon) \,=\, \log \sigma_n \,-\, \log \sigma_{n-1} \,-\,\left[ n f_0(\lambda n)\,-\, (n-1)f_0(\lambda (n-1))\,+\, 0.5 \log \frac{n}{n-1}\right]\,.
\label{eq:mainSSB}
\]
The main point is that the expression in square brackets is known as it is dictated by the known function $f_0(\lambda n)$ in \eqref{f0SSB}.
(In addition, the constant $c_0=0.5$ is fitted from the data, and it results in a small correction numerically.)
Equation \eqref{eq:mainSSB} is our main tool for computing the holy grail function in the model with SSB from the ratios of cross-section
data for $n=7$ and $n=6$ particles in the final state.

 \begin{figure}[t]
\begin{center}
\begin{tabular}{cc}
\hspace{-1.4cm}
\includegraphics[width=0.55\textwidth]{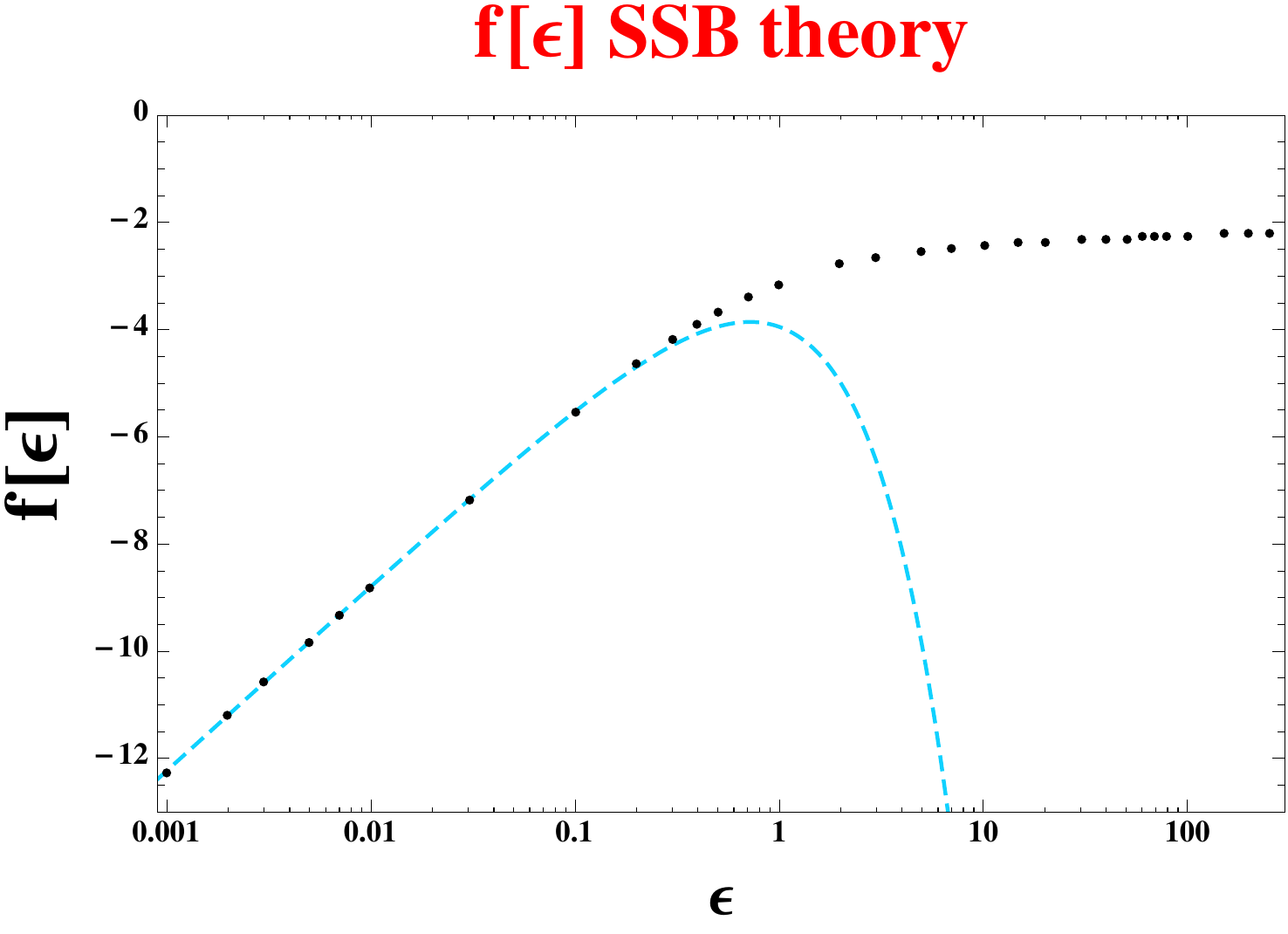}
&
\includegraphics[width=0.55\textwidth]{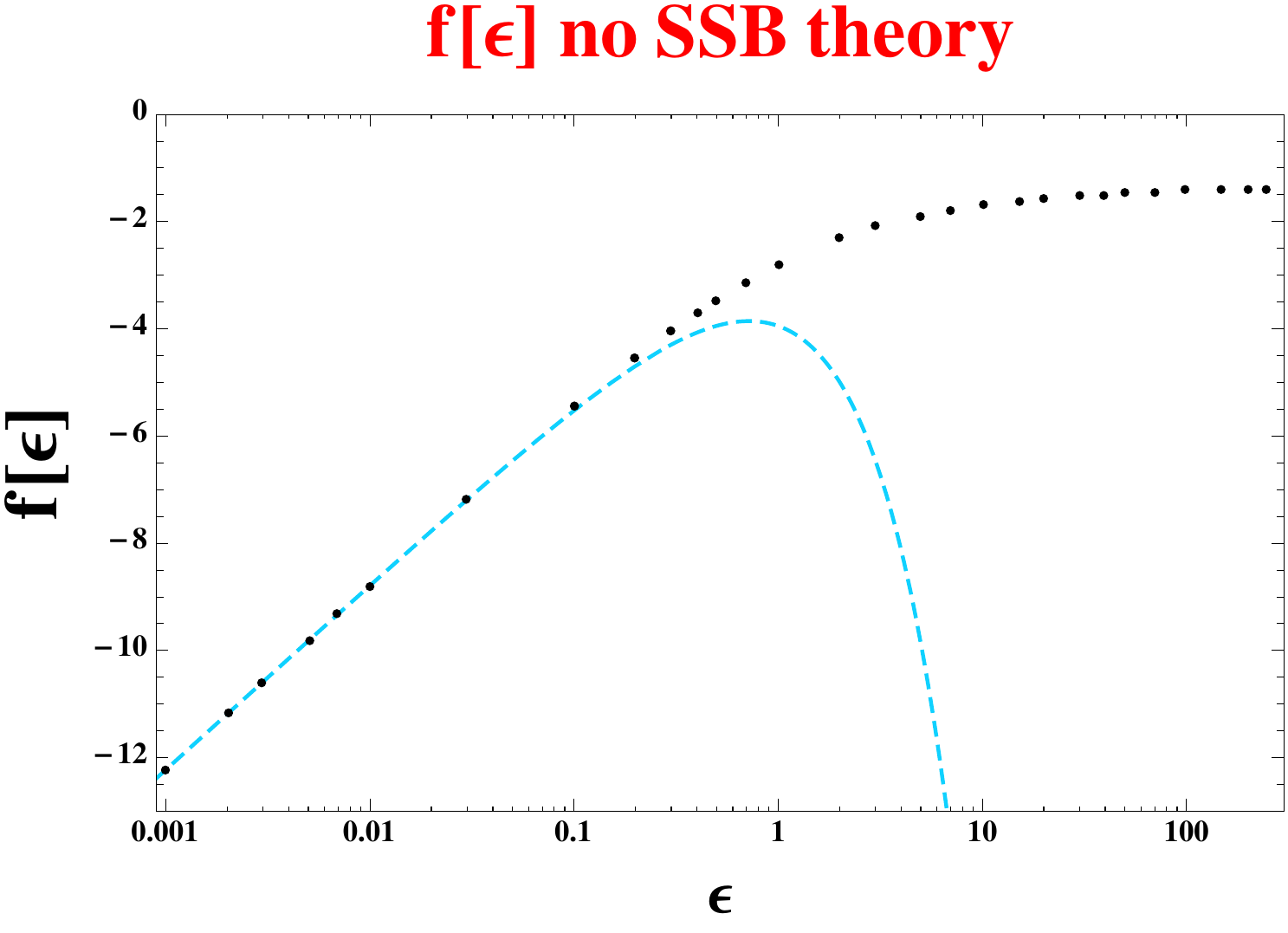}
\\
\end{tabular}
\end{center}
\vskip-.6cm
\caption{
Plots of $f(\varepsilon)$ extracted from the $\log \sigma_7^{\rm tree} /\sigma_6^{\rm tree}$ MadGraph data 
in the SSB model, and the $\log \sigma_7^{\rm tree}/ \sigma_5^{\rm tree}$ realisation of  $f(\varepsilon)$
in the model without SSB. The results perfectly match $f(\varepsilon)_{\rm asympt}$
for $\varepsilon < 1$ depicted in light blue.
}
\label{fig:f3}
\end{figure}
 \begin{figure}[h!]
\begin{center}
\begin{tabular}{cc}
\hspace{-1.4cm}
\includegraphics[width=0.55\textwidth]{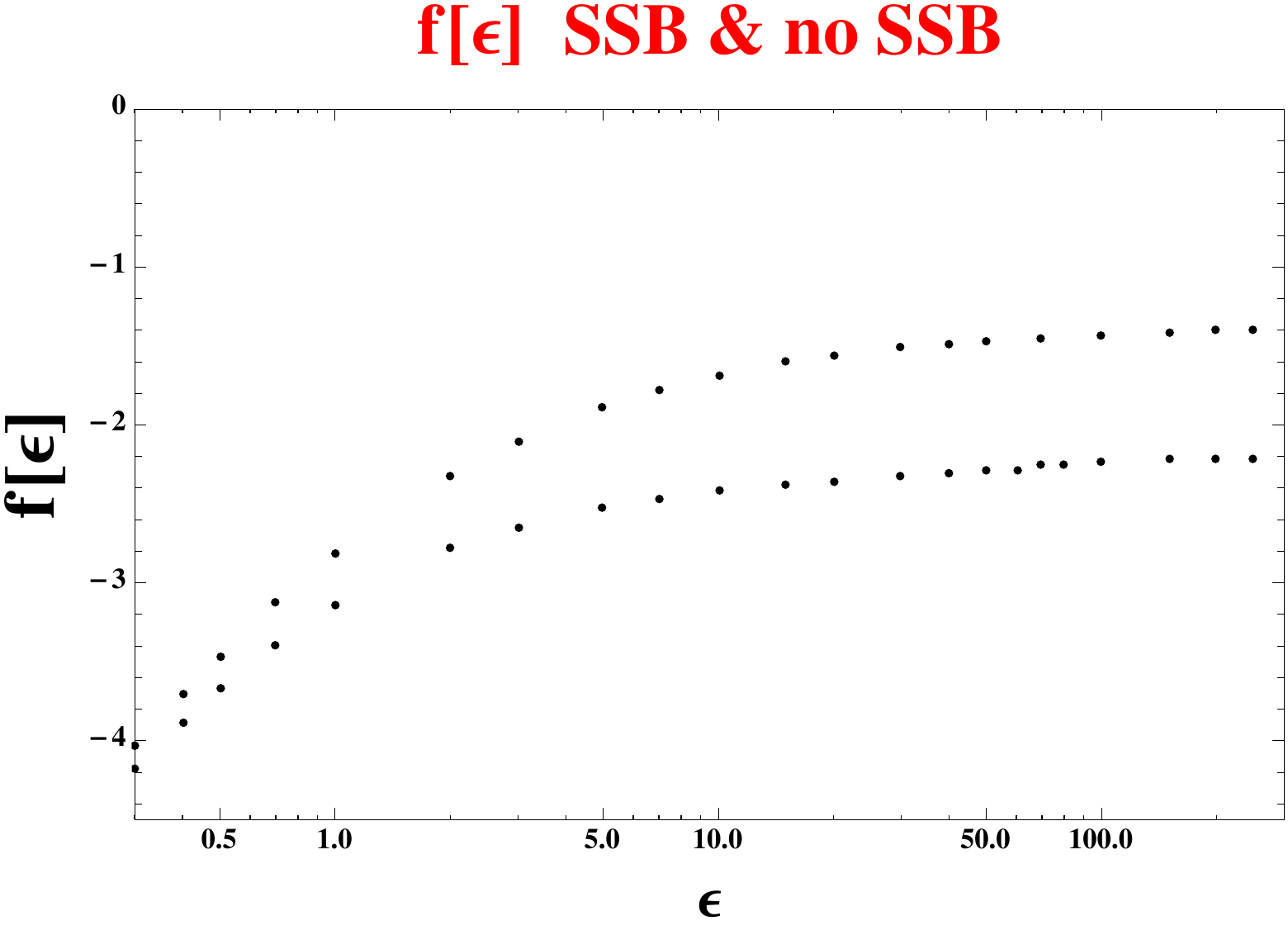}
&
\includegraphics[width=0.55\textwidth]{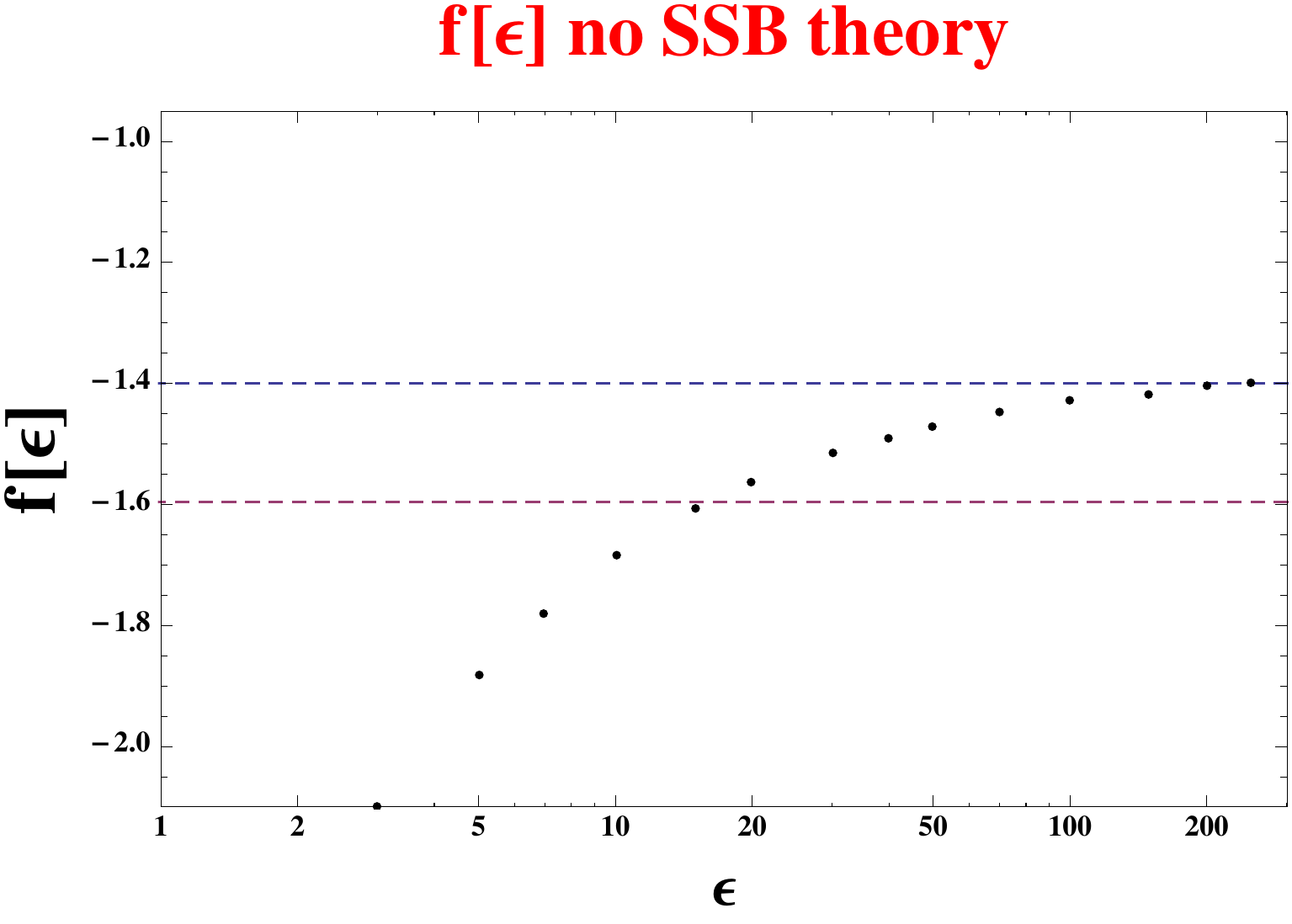}
\\
\end{tabular}
\end{center}
\vskip-.6cm
\caption{
Plots of $f(\varepsilon)$ in the broken and unbroken theory for medium toi large values of $\varepsilon$. In the UV regime 
the functions asymptote to 
$f(\varepsilon=250)^{\rm SSB} \simeq -2.2$ and $f(\varepsilon=250)^{\rm no\, SSB} \simeq -1.4$. 
The plot on the right depicts $f(\varepsilon)^{\rm no\, SSB}$ and shows that it exceeds the asymptotic lower limit $-\log \pi^2/2\simeq -1.6$
obtained from the O(4) symmetric classical solution \cite{Son,Bezrukov:1995qh,LRT}.}
\label{fig:f4}
\end{figure}

The scattering amplitude into the final state with  $n=6$ bosons 
in the model \eqref{eq:LSSB} contains 2,485 Feynman diagrams at tree level. This is still a large 
enough number of diagrams (to be in the regime of a `high-order' perturbation theory), so  we can use the improved large-$n$ subtraction formula \eqref{eq:mainSSB}.
The cross-sections $\sigma_6^{\rm SSB}$ are computed on the same $\varepsilon$ grid as before.
The characteristic value at $\varepsilon=1$  is now
$\sigma_6^{\rm SSB}=\, 1.77 \times 10^{-9} \, {\rm pb}$, and at
$\varepsilon=30$  the rate is 
$1.649 \times 10^{-9} \, {\rm pb}$. 

Our results for the function $f(\varepsilon)$ in the Higgs theory \eqref{eq:LSSB} derived form the numerical 
cross-section data using \eqref{eq:mainSSB} with $n=7$,
are shown on the left plot in Fig.~\ref{fig:f3}. This plot also shows a perfect match to the known $f(\varepsilon)_{\rm asympt}$
expression \eqref{feSSB} at $\varepsilon < 1$, which is shown as a dashed curve in light blue.
As another test of self-consistency of our procedure, we have checked that $f(\varepsilon)$ obtained from the $7-6$ computation
in fact matches closely the function extracted from a similar $6-5$ computation.

For completeness, and to compare with the numerical predictions based on the semi-classical analysis in \cite{LRT,Son,Bezrukov:1995qh},
we have also computed $f(\varepsilon)$ in the unbroken theory \eqref{eq:LnoSSB}. Our results for 
for $\sigma_7^{\rm n0\, SSB}$ for $0.1 \le \varepsilon\le 250$ are shown in Fig.~\ref{fig:f2} as the lower contour (in purple).
The two values of the cross-sections in the two models appear to converge in the UV. This is not surprising, since at very high energies,
all mass parameters become irrelevant and there is little difference between the models with the `right' and the `wrong' sign of the mass-squared term.

To determine $f(\varepsilon)$ in the unbroken theory from the diagrammatic computation, we use the master formula,
\[
2\, f(\varepsilon) \,=\, \log \sigma_n \,-\, \log \sigma_{n-2} \,-\,\left[ n f_0(\lambda n)\,-\, (n-2)f_0(\lambda (n-2))\,+\, 0.5 \log \frac{n}{n-2}\right]\,,
\label{eq:mainnoSSB}
\]
for the $n=7$ and $n=5$ rates (in the unbroken theory there are no 3-point vertices and the amplitudes are non-vanishing only for odd values of $n$).
Our results are shown on the right plot in Fig.~\ref{fig:f3}. 
The left panel in Fig.~\ref{fig:f4} plots the results for $f(\varepsilon)$ functions in the SSB model and the unbroken theory side by side for moderate
to large values of $\varepsilon$. 

In principle we should keep in mind that our analysis is based on the applicability of the
subtraction formulae which assume that $n$ is large enough to ensure that $1/n$ corrections are negligible in the subtraction formulae. 
Thus our derivation of $f(\varepsilon) $ in the unbroken theory, which is based on the $n=7$ and $n=5$ data with (280 and 10 Feynman diagrams)
is 
less robust in comparison to our main SSB theory results based on the $n=7$ and $n=6$ data with 34,330 and 2,485 Feynman diagrams.
However computations of $2\to 9$ processes with MadGraph, which would be the next step in the unbroken theory, is beyond the scope of this paper.

\subsection{Multi-particle cross-sections}

\begin{figure}[t]
\begin{center}
\includegraphics[width=0.8\textwidth]{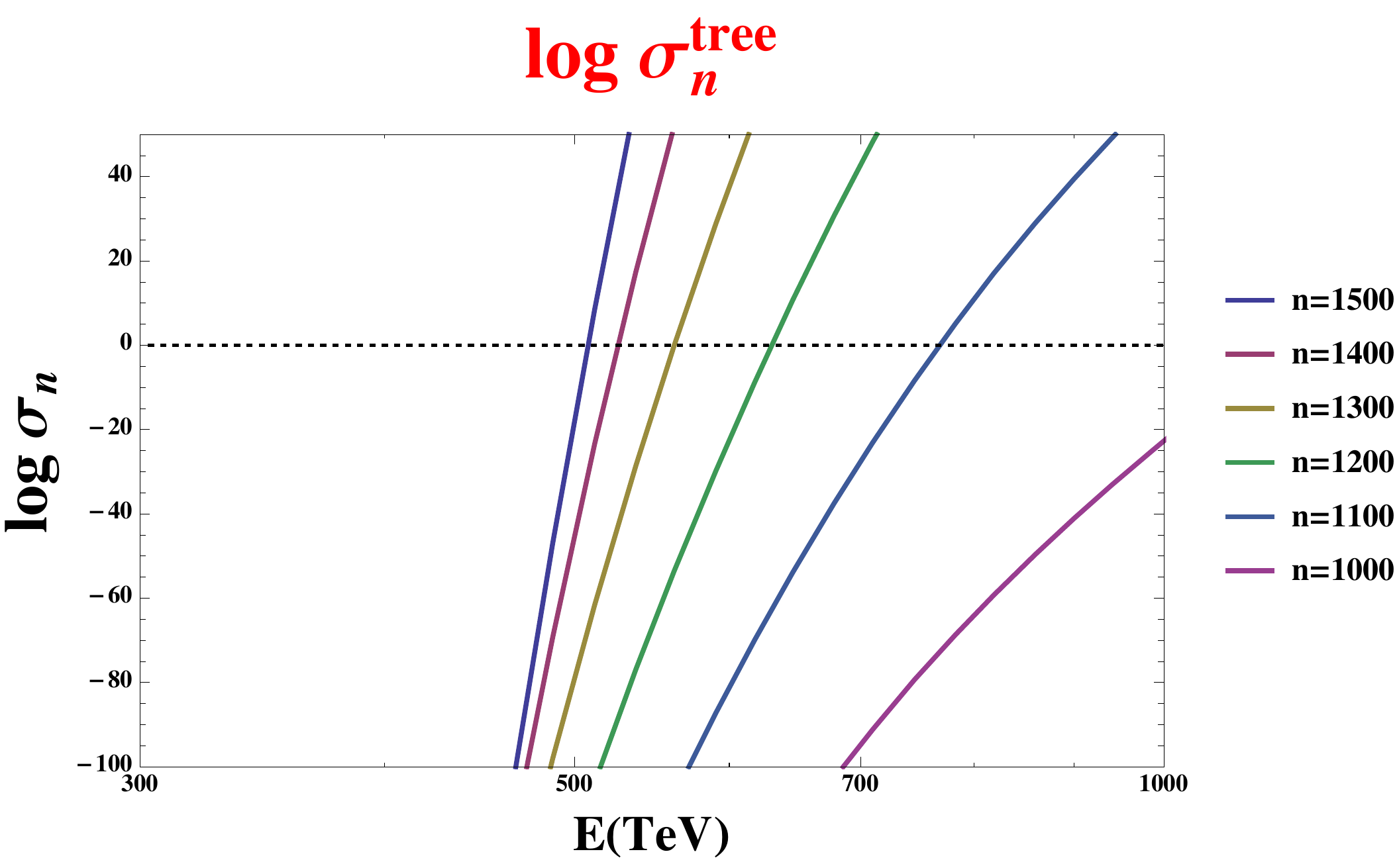}
\end{center}
\vskip-.4cm
\caption{
Plots of multi-particle tree-level cross-sections $\sigma_n^{\rm tree}$ in the scalar model with SSB
as the function of energy $E$ for a range of final-state multiplicities between $n=1000$ and $n=1500$.
}
\label{fig:f5}
\end{figure}

Having determined the $n$-independent kinetic energy function $f(\varepsilon)$ allows to us to compute multi-particle 
cross-sections at any $n$ in the large-$n$ limit. The tree-level multi-particle cross-sections $\sigma_n^{\rm tree}$ in the scalar theory with SSB 
are obtained via
\[
\log \sigma_n (E)\,=\,  n\left(f_0(\lambda n)\,+\, f(\varepsilon)\right)\,,
\label{eq:nFtree2}
\]
with $\varepsilon(E,n)=(E-nM)/(nM)$ and 
\[
f_0(\lambda n)^{\rm SSB\, tree}\,=\,  \log\left(\frac{\lambda n}{4}\right) -1\,, 
\]
and we set $\lambda=1/8$ and $M=125$ GeV. In Figure~\ref{fig:f5} we plot the cross-sections $\sigma_n^{\rm tree}$ in this theory 
as a function of energy $E$ for a range of final-state multiplicities between $n=1000$ and $n=1500$.
The choice of such high values of particles in the final state follows from selecting the regime where the tree-level cross-sections become unsuppressed.
This occurs when the positive $f_0(\lambda n)^{\rm SSB\, tree}$ factor is able to compensate the negative values of $f(\varepsilon)$. As the result we see that perturbative cross-sections grow very steeply with energy, and the interesting range of energies where the $\log \sigma_n^{\rm tree}$ crosses zero occurs is the $E\sim 500$ TeV regime. At these energies 
the tree-level cross-sections grow exponentially violating perturbative unitarity. The energy regime where this happens in Fig.~\ref{fig:f5} 
is in agreement with the estimates obtained in \cite{JK}.
What is interesting, is that the energy scales
where perturbation theory breaks down (judging from the leading tree-level analysis here) occurs at energies only a (few)$\times 10^1$  above what could be directly tested experimentally with a hadron FCC collider. 

 \begin{figure}[h!]
\begin{center}
\includegraphics[width=0.8\textwidth]{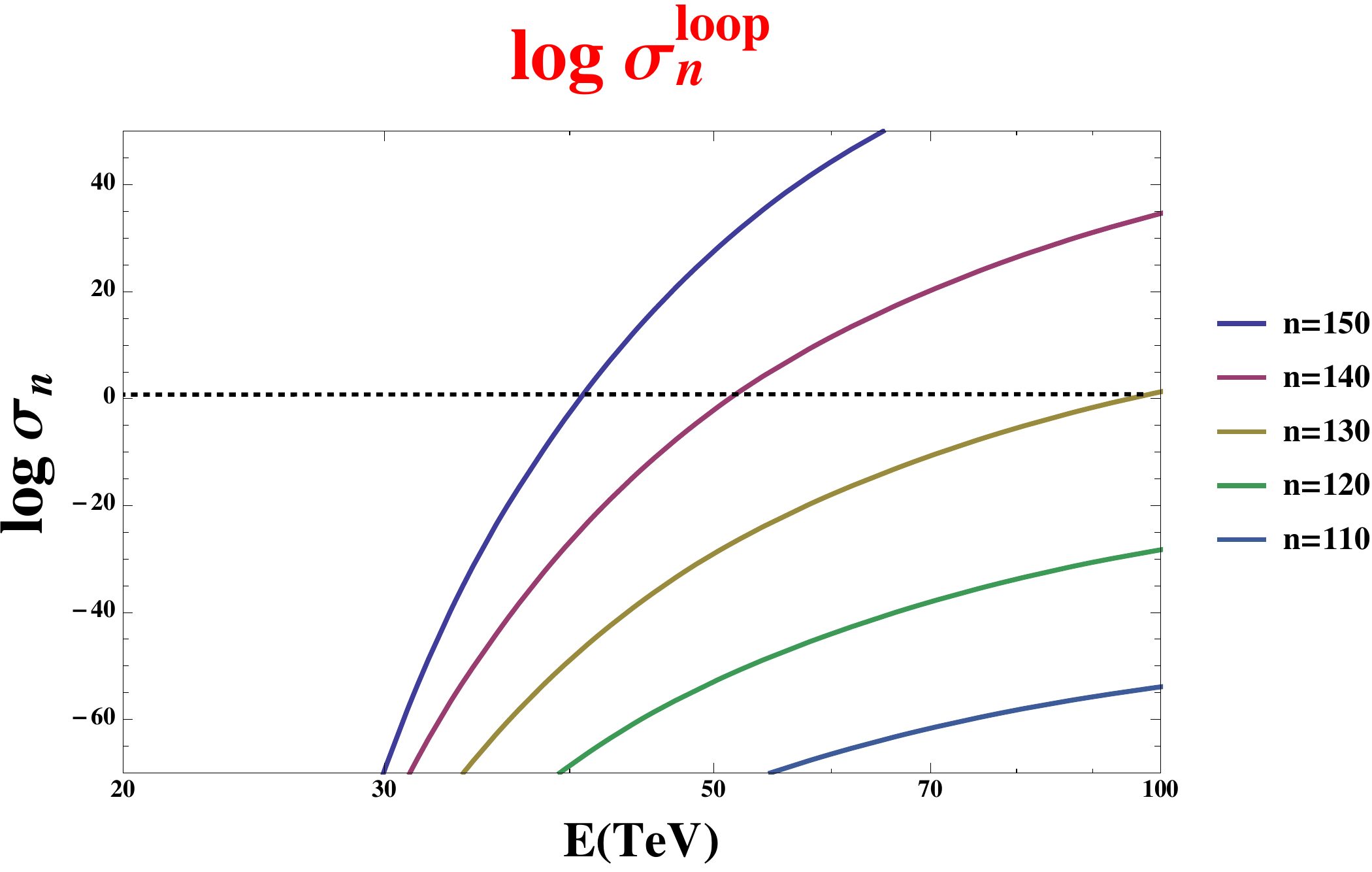}
\end{center}
\vskip-.4cm
\caption{
Results for multi-particle cross-sections $\sigma_n^{\rm loop}$ with the leading-loop-resummation factor \eqref{eq:nFloop}-\eqref{f0SSBloop}
and the single Higgs production form-facror in the model with SSB \eqref{eq:LSSB}. The logarithm of the cross-section \eqref{eq:sigmaff} is plotted
as the function of energy for a range of final-state multiplicities between $n=110$ and $n=150$.
}
\label{fig:f6}
\end{figure}

Let us now consider the effect of loop corrections.
The 1-loop corrected multi-particle amplitudes on multi-particle thresholds are known \cite{Voloshin83,Smith84}, and the result 
in the broken scalar theory \eqref{eq:LSSB} is given by \cite{Smith84},
\[
{\rm SSB}: \quad
{\cal A}_{1^*\to n}^{\rm tree + 1 loop}\,=\, n!\, (2v)^{1-n} \left(1+n(n-1)\frac{\sqrt{3} \lambda}{8\pi}\right)
\,.
\label{eq:amplnh1}
\]
It was shown in Ref.~\cite{LRST}, based on the analysis of leading singularities of the multi-loop expansion around singular generating functions in scalar field theory,
that the 1-loop correction exponentiates,
\[
{\cal A}_{1^*\to n}^{\rm loops}\,=\,{\cal A}_{1^*\to n}^{\rm tree}\times\, \exp\left[B\, \lambda n^2\,+\, {\cal O}(\lambda n)\right]\,
\label{expB} 
\]
in the limit $\lambda \to 0,$ $n\to \infty$ with $\lambda n^2$ fixed, where $B$ is the constant factor determined from the 1-loop calculation,
\begin{eqnarray}
{\rm model\,with\, SSB~\eqref{eq:LSSB}}: \quad
B &=& +\, \frac{\sqrt{3}}{8 \pi} \,, \label{BSSB}\,,\\
{\rm unbroken\, model~\eqref{eq:LnoSSB}}: \quad
B &=& -\, \frac{1}{64 \pi^2}\left(\log(7+4\sqrt{3})-i\pi\right) \,. \label{BnoSSB} 
\end{eqnarray}
As the result, the the leading-order multi-loop exponentiation leads to the the exponential enhancement of the multi-particle cross-section
in the Higgs model, {\it cf.} Eqs.~\eqref{f0SSB}, \eqref{f0SSBloop},
\[
f_0(\lambda n)^{\rm loop} \,=\, \log\left(\frac{\lambda n}{4}\right) -1 \,+\, 2B\,\lambda n\,.
\]

Finally we can also include the single Higgs production form-factor in front of the exponential factor in the cross-section,
to correct for our use of the effective Higgs-gluon vertex in the large energy limit,\footnote{I would like to thank Michael Spira for 
this suggestion.}
\[
\sigma_n^{\rm loop} \,=\, \left(m_t/E\right)^4 \log^4 (m_t/E)^2\,e^{n\left(f_0(\lambda n)\,+\, f(\varepsilon)\right)}\,,
\label{eq:sigmaff}
\]
where $m_t$ is the top mass.
In total we have
\begin{eqnarray}
\log \sigma_n^{\rm loop} &=& n\left(f_0(\lambda n)^{\rm loop}\,+\, f(\varepsilon)\right)\,-\, 4\left(\log \left(E/m_t\right) -\log \log (E^2/m_t^2)\right)\,,
\label{eq:nFloop}
\\
\label{f0SSBloop}
f_0(\lambda n)^{\rm loop} &=&  \log\left(\frac{\lambda n}{4}\right) -1 \,+\, \sqrt{3}\, \frac {\lambda n}{4\pi}\,, 
\end{eqnarray}
where the last equation is consistent with \eqref{BSSB}
and leads to the exponential enhancement of the cross-section $\sigma_n$,
at least in the leading order in $n^2 \lambda$. 
The form-factor correction -- the last term on the right hand side of \eqref{eq:nFloop} -- 
grows with $n$ only logarithmically\footnote{At large $n$ and large $\varepsilon$ limit it is
$\simeq -4\log(n\varepsilon) + 4 \log \log (n\varepsilon) + 4 \log \left( (m_t/M) \log (M/m_t)\right).$ }
compared to the linear in $n$ terms in the first term. (At $E=50$ TeV the form-factor gives the correction
 $-\, 4\left(\log \left(E/m_t\right) -\log \log (E^2/m_t^2)\right) \simeq -12.95$ in the exponent.)

Our results for $\sigma_n^{\rm loop}$ including the form-factor and the exponentiated loop factor (the last term in \eqref{f0SSBloop})
for the Higgs model \eqref{eq:LSSB} are shown in Fig.~\ref{fig:f6}
for a range of final-state multiplicities between $n=110$ and $n=150$.
We can see that the loop-enhancement has reduced the energy scale (and multiplicities) by a factor of 10,
and the scale for the perturbation theory breakdown or equivalently the scale of new phenomena to set in
is now reduced to 40-50 TeV which is pretty much within the energy reach of the 100 TeV FCC collider,
in agreement with the estimate in \cite{JK}.

Of course, one should keep in mind that the set-up in Eqs.~\eqref{eq:nFloop}-\eqref{f0SSBloop} is merely an optimistic
phenomenological model. In general 
the even higher-order  effects of loop exponentiation will be present such that,
\[
f_0(\lambda n)^{\rm all \,loops} \,=\,  \log\left(\frac{\lambda n}{4}\right) -1 \,+\, \sqrt{3}\, \frac {\lambda n}{4\pi}
 \,+\, {\rm const} \left( \frac {\lambda n}{4\pi}\right)^2  \,+\, {\rm const'} \left( \frac {\lambda n}{4\pi}\right)^3 \,+\, \ldots
\,, 
\]
and can change the cross-sections contours in Fig.~\ref{fig:f6}. (Note that the value of the loop expansion parameter
$\frac{\lambda n}{4\pi}$ are $\simeq 1$ for $n=100$ and $\simeq 1.4$ for $n=140$.)

 \section{Conclusions}

 Our diagrammatic approach is conceptually different (but also complimentary) to the 
 semi-classical considerations followed in the earlier literature.
 The exponential form of the cross-section in the large-$n$ limit, 
 \[\sigma_n \,\sim\, \exp \left[ \frac{1}{\lambda}\, \lambda n F(\lambda n, \epsilon)\right]\,:=\,
 \exp \left[ \frac{1}{\lambda}\, {\cal F}(\lambda n, \epsilon)\right]
 \,,
 \]
 is strongly suggestive of an underlying semi-classical origin of the multi-particle cross-section.
 In particular, there is a strong similarity between the purely perturbative multi-particle processes considered here and the 
 $B+L$-violating non-perturbative reactions in the instanton sector of the Standard Model discussed originally in \cite{R}.

The idea that semi-classical methods can be also used in the perturbative sector of the theory
was put forward and explored by a number of authors including 
 Refs.~\cite{Son,Bezrukov:1995qh,LRT,Voloshin:1990mz,Gorsky:1993ix}.
At tree level the holy grail function $F$ or ${\cal F}$ can indeed be reconstructed numerically if one can determine
certain singular classical solutions to the boundary value problem \cite{Son}, as explained in \cite{LRT,Son,Bezrukov:1995qh}.
In practice this procedure was carried out in the case of the unbroken $\phi^4$ theory \eqref{eq:LnoSSB} and
based on finding numerically the singular solutions with the hypothesised O(4) symmetry. In this approach a lower bound on the tree-level cross-section \eqref{eq:hg} was derived in \cite{Bezrukov:1995qh, LRT} which corresponds to an upper bound on 
the absolute value of $|f(\varepsilon)|$. In particular it was found that at infinite energies, $\varepsilon \to \infty$ 
the function $f(\varepsilon) \to -\log(\pi^2/2)\simeq -1.6$. In our case, the asymptotic value appears to be smaller
in magnitude in the non-SSB theory, $f(\varepsilon) \to -1.6$, as can be seen from the right panel in Fig.~\ref{fig:f4}.
The fact that  $f(\varepsilon) \to -\,|{\rm const}|$ implies,
\[\sigma_n^{\rm tree\, noSSB} \,\gtrsim \, e^{-\,|{\rm const}|\, n} \, e^{n f_0(\lambda n)}  \qquad {\rm at} \,\, E \to \infty\,.
\label{eq:O4bound}
\]
An often quoted misreading of this result is the statement that perturbative cross-sections remain unobservable 
in the multi-particle limit, even at infinitely high energies,  due to a rising with $n$ exponential suppression factor  $e^{-\,|{\rm const}|\, n}$.
This, of course, is not the case as the plots in Figs.~\ref{fig:f5} and \ref{fig:f6} demonstrate: the growing function $e^{n f_0(\lambda n)} $ 
compensates the suppression in $e^{-\,|{\rm const}|\, n}$ for any $\varepsilon$
already at moderately high values of $\lambda n$. 

\medskip
  
One advantage of the diagrammatic approach followed in this paper is its simplicity, and also the fact that one should be able to 
apply it in any theory, ultimately including the full Gauge-Higgs theory of the Standard Model weak sector by generalising the
non-relativistic results of \cite{VVK1,VVK2} to the general-$\varepsilon$ case. 
We leave this to future work.

\medskip
Since in our case the calculations are carried out within the first principles perturbative approach, we also know that
as soon as the regime is reached where the theory breaks down and violates unitarity, this implies that 
we really are falsifying the perturbative technique itself, and not a bound arising from a semi-classical treatment.
The perturbation theory break-down found here occurs in two cases: a) within the tree-level approximation in the energy-multiplicity regime of 
Fig.~\ref{fig:f5}, and b) within the leading-order in the loop expansion approximation in the regime corresponding to
Fig.~\ref{fig:f6}.

The main technical challenge which still needs to be addressed is how to account for all the remaining higher-loop corrections, relevant in the
regime $\lambda n \sim 1$.
Even the leading-order exponentiation of the loop corrections result (the last term on the right hand side of \eqref{eq:nFloop}) 
which was essential for lowering the characteristic energy scale from Fig.~\ref{fig:f5} to Fig.~\ref{fig:f6} by an order of magnitude, has been
derived only in the multiparticle threshold limit; the fool $\varepsilon$-dependence of leading loop corrections remains unknown.

Another (perhaps less crucial) technical limitation of our simple derivation is that we have concentrated only on sub-processes with
a single virtual Higgs in the s-channel in \eqref{eq:singleh}. We have not considered here the effect of possible numerical partial cancellations 
between the s-channel ${\bar t} \to h^* \to hh$ and box diagram processes ${\bar t} \to hh$ and  in the double Higgs case and generalisations  
for the multi-Higgs case, see e.g. the discussion in \cite{Li:2013rra}. However, in the absence of the symmetry reason, we do not expect that such
partial cancellations could significantly modify the exponential growth of the $s$-channel processes \eqref{eq:singleh}.
This conclusion is also n agreement with the discussion in section 4 of \cite{JK} where we have seen that the exponential growth persists
in the similar case of the weak vector boson fusion, $VV \to h^* \to hh \to n\times h$ vs $VV \to h^*h^*\to n\times h.$

\bigskip

We have shown that in very high energy scattering events, perturbative rates for production of multiple Higgs bosons grow with increasing energy, eventually violating perturbative unitarity and resulting in the breakdown of the ordinary weakly-coupled perturbation theory.
The energy scales where electro-weak processes can enter this regime are potentially within the reach of the 100 TeV future hadron colliders, 
or at least not much above it.
It was argued in \cite{JK} that novel physics phenomena must set in before these energies are reached: either the electroweak sector becomes  non-perturbative in this regime, or additional physics beyond the Standard Model might be needed. 

\section*{Acknowledgements}

I am grateful to Joerg Jaeckel, Michelangelo Mangano, Gavin Salam and Michael Spira for discussions and comments about computing multi-particle rates 
from first principles at finite $n$, and to Olivier Mattelaer and Gunnar Ro for sharing with me their expertise with MadGraph.
This work is supported by the STFC through the IPPP grant, 
and by the  Royal Society Wolfson Research Merit Award.
%

\bigskip

\bibliographystyle{h-physrev5}

\end{document}